\documentclass[epjCONF,columns,final]{svjour}
\usepackage{graphics}
\usepackage[varg]{txfonts} 
\usepackage[latin1]{inputenc}
\newcommand{\etal}{{\em et al.}}

\newcommand{\sqrtsnn}{$\sqrt{s_{\rm NN}}$}
\newcommand{\vtwo}{${\rm v}_2$}
\newcommand{\shear}{$\eta/\mathrm{s}$}

\newcommand{\pp}{pp}

\newcommand{\pbpb}{Pb--Pb}
\newcommand{\auau}{Au--Au}

\newcommand {\gevc} {\mbox{\rm GeV$\kern-0.15em /\kern-0.12em c$}}
\newcommand {\tevc} {\mbox{\rm TeV$\kern-0.15em /\kern-0.12em c$}}

\newcommand {\tev} {\mbox{${\rm TeV}$}}

\newcommand {\mev} {\mbox{${\rm MeV}$}}

\session-title{Heavy Ion (II) - soft}
\begin{document}
\title{Soft physics in \pbpb{} collisions at the LHC}
\author{P.G.~Kuijer\inst{1}\fnmsep\thanks{\email{Paul.Kuijer@nikhef.nl}}}
\institute{Nikhef, National institute for subatomic physics, P.O. Box 41882, 1009 DB Amsterdam, Netherlands}
\abstract{
The LHC results on particle and transverse energy production and Bose-Einstein correlations show that the system produced in Pb-Pb collisions at $\sqrt{s_{\rm NN}}=2.76 {\rm TeV}$ is significantly larger, lives longer and is hotter and denser than at RHIC energies. The particle spectra and soft particle correlations allow a more detailed study of the properties of the produced system and its initial conditions. This contribution highlights some of the first soft physics results from the LHC. 
} 
\maketitle
\section{Introduction}
\label{intro}
Nucleus--nucleus collisions differ significantly from \pp{} collisions because the large volume and large amount of energy available allow rescattering and additional production of the gluons and quarks after the initial hard scattering stage of the collision. If the system is sufficiently large and the energy density sufficiently high the quarks and gluons can thermalize quickly and form a quark-gluon plasma~(QGP). The quark gluon plasma is the only state accessible in the laboratory in which quarks and gluons are deconfined. After the formation of a QGP the system expands and cools down. Given a sufficiently large system the expansion can be described by hydrodynamic models. When the temperature becomes lower a phase transition to an hadron gas takes place. In order to reduce the contribution of the pre-equilibrium and hadronic phases to the final state the size and lifetime of the QGP should be maximized. Experimentally this can be achieved by increasing the beam energy. 

It has been established at RHIC that a hot and dense medium of strongly interacting constituents is formed in \auau{} collisions at \sqrtsnn{}=0.2~\tev{} and that this medium is best described as a quark-gluon plasma~\cite{ref:rhic}. At RHIC, the surprising conclusion was that the QGP behaves as a liquid with a rather small shear viscosity.

In November 2010 the LHC provided \pbpb{} collisions with an energy of \sqrtsnn{}=2.76~\tev{}, about 14~times larger than the energy achieved at RHIC. At this higher energy the produced medium is expected to have a higher initial temperature and density and therefore to expand to larger volumes as well as to exist longer. The larger number of particles available in the final state allows a more precise experimental study of the \pbpb{} collisions.

The first LHC results have indeed confirmed that a larger and hotter system has been produced and that the shear viscosity of the QGP is low. The LHC experiments have now started to provide more insight into the soft processes involved in the production and evolution of the QGP. This contribution describes the first results obtained by the LHC experiments.
\section{Particle production and system size}
\label{sec:production}
\begin{figure}
\resizebox{0.9\columnwidth}{!}{%
\includegraphics{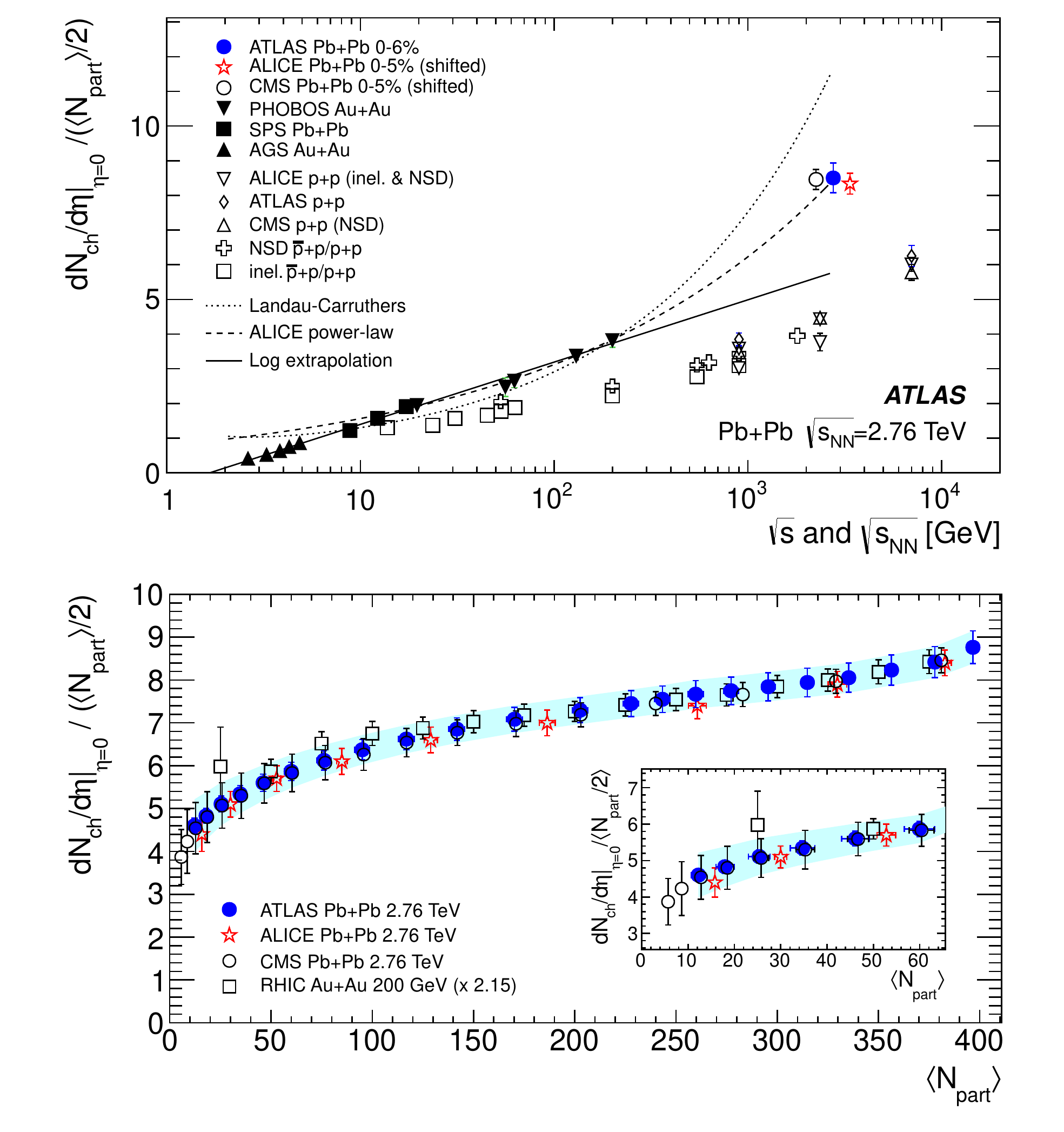}%
}
\caption{
Colour online. Top: Charged particle multiplicity density at mid-rapidity per participant pair for various collision systems and energies. The lines represent various phenomenological extrapolations from the SPS and RHIC energies to the LHC. Bottom: Centrality dependence of the multiplicity density per colliding nucleon pair. The RHIC results are scaled up by a factor 2.15. Figures taken from~\cite{ref:mult}.
}%
\label{fig:mult}%
\end{figure}

The charged particle multiplicity density for central \pbpb{} collisions at the LHC at 2.76~\tev{} is ${\rm d}N_{\rm ch}\approx{}1600$, see~Fig.~\ref{fig:mult}. This is a factor~2.15 larger than at top RHIC energies. Compared to \pp{} collisions at the same energy the particle production is increased by a factor~1.9. The growth with energy is faster in \pbpb{} than in \pp{} collisions. 
The bottom panel of Fig.~\ref{fig:mult} shows the centrality dependence of the multiplicity density per participant pair. In this figure the RHIC results are scaled by a factor~2.15, showing that the shape of this distribution does not change much when the collision energy is increased from 0.2~\tev{} to 2.76~\tev{}. The multiplicity density per participant pair increases slowly from about 4 to about 9 from peripheral to central events. As expected the particle production in peripheral AA~events approaches the value in \pp{} collisions. 

The combination of the momentum spectra and multiplicity measurements provides an estimate of the energy density through the Bjorken formula~\cite{ref:bjorken}. The energy density obtained is ${\rm d}E_{\rm t}/{\rm d}\eta\approx2~\tev{}$ for central \pbpb{} collisions~\cite{ref:cmsedens}, which is an increase of more than a factor~2.5 with respect to the energy densities reached at RHIC. 

\begin{figure}
\resizebox{0.9\columnwidth}{!}{%
\includegraphics{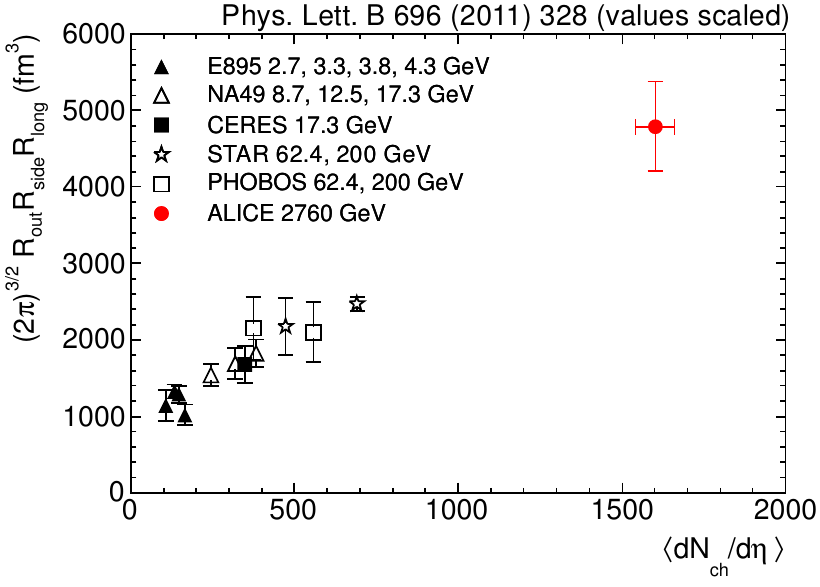}%
}
\resizebox{0.9\columnwidth}{!}{%
\includegraphics{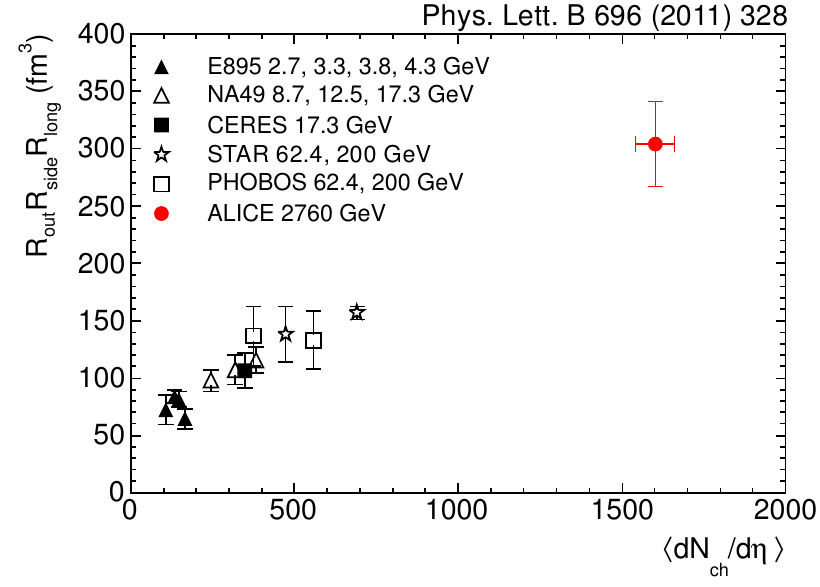}%
}
\caption{
Colour online. 
HBT radii as a function of the charged particle density at mid-rapidity determined at different collision energies. For the charged particle densities reached at the LHC the volume (top) is approximately two times larger than at RHIC energies while the lifetime (bottom) is approximately 40\% larger. Figures taken from~\cite{ref:hbt}.
}
\label{fig:hbt}       
\end{figure}

The volume and the lifetime of the system were deduced from the momentum correlations of identical pions. The measurement provides three length scales, the product of which is an estimate of the size of the homogeneity region, and decoupling time. The resulting volume is shown in the top panel of Fig.~\ref{fig:hbt}, the lifetime in the bottom panel of Fig.~\ref{fig:hbt}. 
At the LHC the volume of the system is a factor two larger than that produced at RHIC. The lifetime increases by more than 30\%. These increases are consistent with hydrodynamical model calculations for LHC energies using parameters tuned to reproduce the RHIC data.
\section{Collective effects}
\label{sec:collective}

\begin{figure}
\resizebox{0.9\columnwidth}{!}{%
\includegraphics{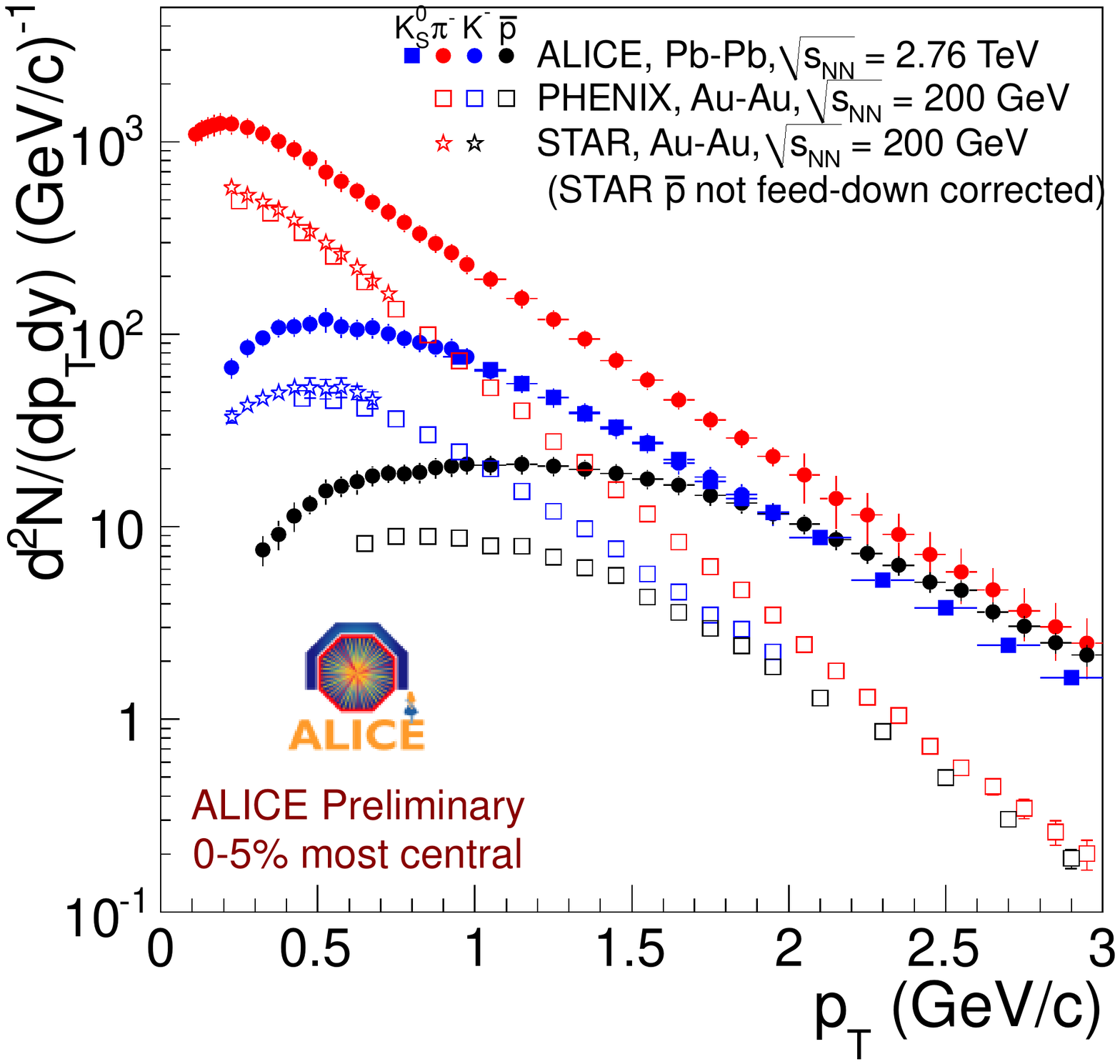}%
}
\resizebox{0.9\columnwidth}{!}{%
\includegraphics{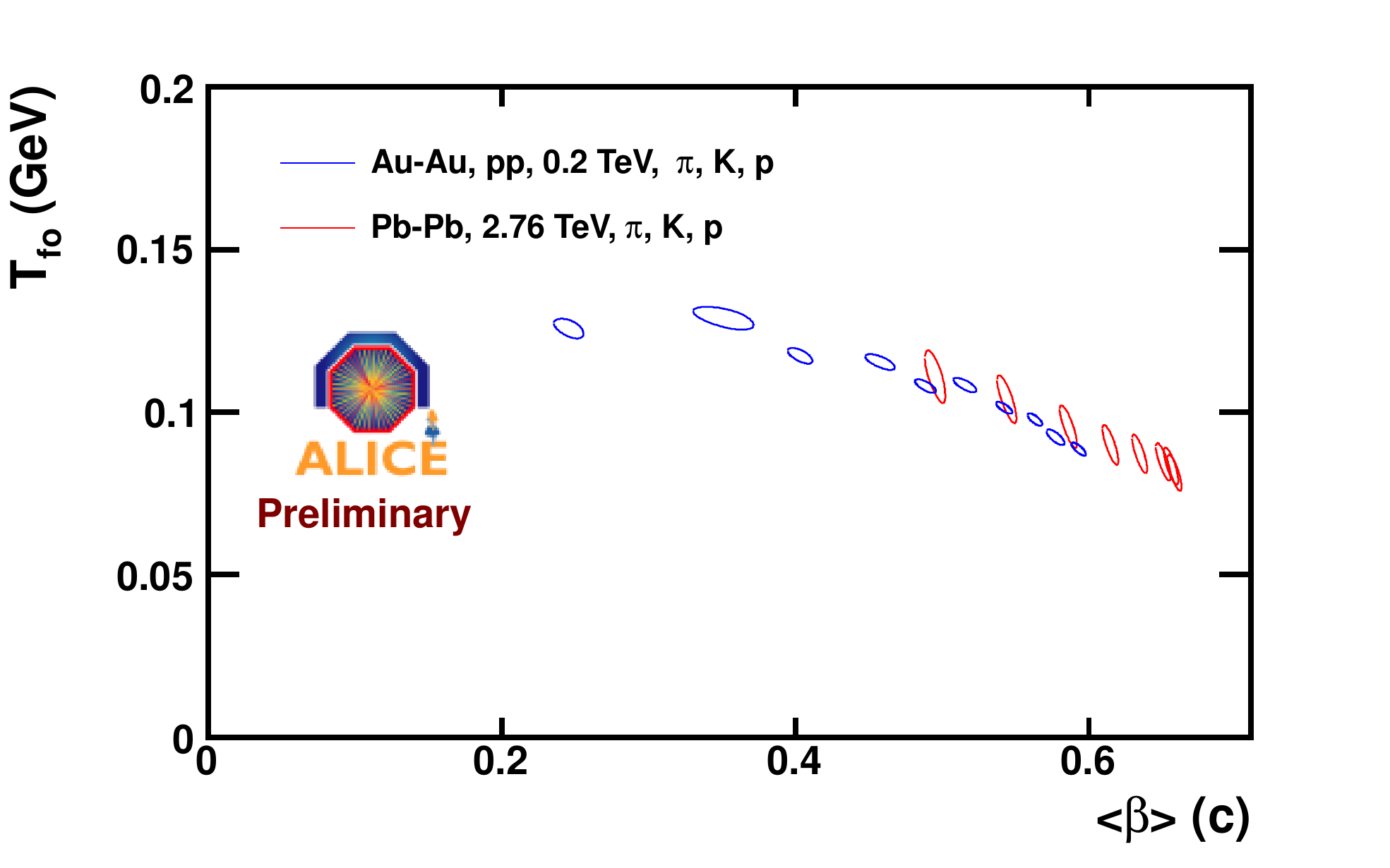}%
}
\caption{
Colour online.
Top: Transverse momentum spectra for charged and neutral kaons, pions and antiprotons for central events for RHIC and LHC energies. Bottom: Freeze--out temperatures and radial flow velocities obtained from blastwave fits to the transverse momentum spectra of identified particles measured at RHIC and LHC. Figure taken from~\cite{ref:blastwave}.
}
\label{fig:blastwave}       
\end{figure}

The hydrodynamic expansion of the QGP shifts the velocities of all particles towards higher values, resulting in harder momentum spectra. Figure~\ref{fig:blastwave} shows a comparison between the transverse momentum spectra of identified particles at RHIC and at LHC energies. The slopes change dramatically with energy, reflecting the increase in radial flow. The values obtained for temperature and radial flow velocities in the final state, i.e. after kinetic freeze out, obtained from a blastwave fit to the spectra~\cite{ref:blastwave} are shown in Fig.~\ref{fig:blastwave} for the most central collisions at RHIC and LHC. 
Increasing the initial density and temperature of the system results in lower freeze--out temperatures and larger radial flow velocities. At the LHC the freeze--out temperature is approximately 80~\mev{} and the radial flow velocity reaches about 60\% of the speed of light. 

\begin{figure}
\resizebox{0.9\columnwidth}{!}{%
\includegraphics{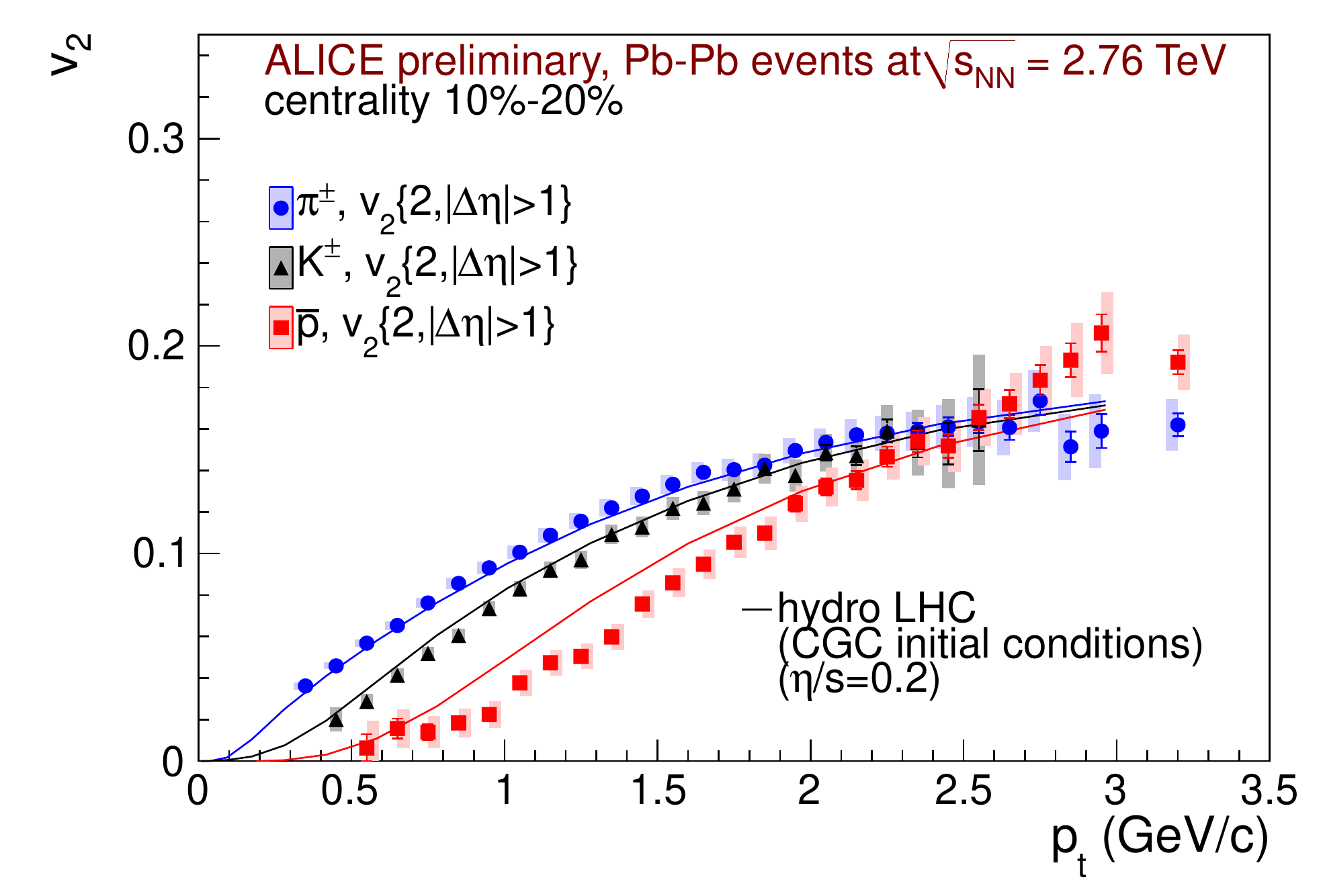}%
}
\caption{
Colour online. 
The elliptic flow coefficient ${\rm v}_2$ for pions, kaons and protons as a function of their transverse momentum. The lines represent calculations including the colour glass initial state and a hydrodynamic expansion with a shear viscosity of $\eta/{\rm s}=0.2$. Figure taken from~\cite{ref:v2pidcent}.
}
\label{fig:v2pidcent}%
\end{figure}
The shape of the overlap area between the colliding nuclei changes from circularly symmetric to an almond shape when going from central collisions to non-central collisions. This anisotropy in coordinate space is transformed into an azimuthal anisotropy in the momentum spectra because the pressure gradients in the QGP are different in the reaction plane as compared to the direction out of the reaction plane. This azimuthal anisotropy is a very characteristic signal of collective behaviour in \pbpb{} collisions. The azimuthal anisotropy can be quantified by the elliptic flow coefficient \vtwo{}. This \vtwo{} corresponds to the second coefficient of the Fourier decomposition of the angular distribution of particles with respect to the reaction plane. Recently also higher order harmonics have been measured. 

The measured \vtwo{} is sensitive to the properties of the QGP as well as to its initial geometry. Studying the dependence of \vtwo{} on the particle type and momentum allows to compare the data to models including both the viscosity and the initial geometry. Figure~\ref{fig:v2pidcent} show the measured \vtwo{} as a function of momentum for pions, kaons and protons. The data show the characteristic mass splitting which is predicted by hydrodynamic calculations. Using the Colour Glass Condensate (CGC) model to describe the gluon density in the incoming nuclei the initial shape and energy density of the produced system can be estimated.  The \vtwo{} calculated from the hydrodynamic expansion of a QGP with a shear viscosy over entropy \shear{}=0.2 qualitatively describes the data. For a full description of all data, additional factors, such as the fluctuations in the initial geometry, will have to be taken into account.

\begin{figure}
\resizebox{0.9\columnwidth}{!}{%
\includegraphics{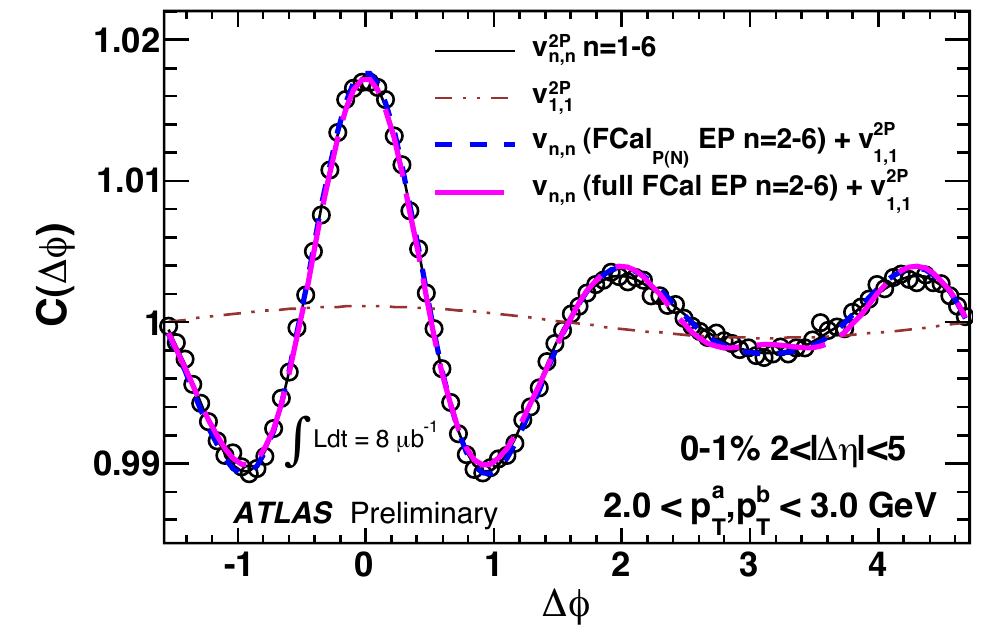}%
}
\caption{
Colour online. 
The measured two particle correlation for the 1\% most central collisions. The solid line represents the angular distribution calculated from the measured flow coefficients. Figure taken from~\cite{ref:ridge}.
}
\label{fig:ridge}%
\end{figure}

Selecting a high momentum particle and measuring the correlation of all other particles as a function of their azimuthal angle difference with respect to the high momentum particle clearly shows jet structures on the near and far side. When the momentum of the selected particle is reduced a characteristic double peak structure appears at the far side, see Fig.~\ref{fig:ridge}. This structure was originally thought to be evidence for the shock wave produced by partons moving through the medium, the Mach cone effect. However, the structure can also be described very well by calculating the correlations assuming only flow. The curve shown in Fig.~\ref{fig:ridge} was calculated from the measured flow coefficients, including the higher order coefficients. In this picture the double peak structure is mainly caused by the interplay between the second and third flow coefficients.
\section{Conclusion}
Quickly after the first \pbpb{} data from \pbpb{} collisions became available the LHC experiments confirmed that the size, lifetime and energy density of the produced system were significantly larger than at RHIC energies. At $\sqrt{s_{\rm NN}}=2.76 TeV$ the energy density is more than 2.5 times larger than at $\sqrt{s_{\rm NN}}=0.2 TeV$. The volume is about two times larger and the lifetime is about 30\% larger. 

Similar to the RHIC results the soft particle azimuthal correlations indicate a QGP with a low viscosity. These results can also describe the double peak structure shown in Fig.~\ref{fig:ridge} which was originally thought to be evidence for the shock wave produced by partons moving through the medium (Mach cone effect).

\end{document}